\documentclass[12pt,letterpaper]{article}

\usepackage{xcolor, colortbl}
\usepackage{amssymb, amsmath, amsthm}
\usepackage{graphicx}
\usepackage{natbib}
\usepackage{float}
\usepackage{setspace}
\usepackage{algorithm}
\usepackage{algorithmic}

\newtheorem{proposition}{Proposition}

\definecolor{lightgray}{gray}{0.9}

\newcommand{\ba}{\mathbf{a}}
\newcommand{\bA}{\mathbf{A}}

\newcommand{\bB}{\mathbf{B}}

\newcommand{\bC}{\mathbf{C}}
\newcommand{\bD}{\mathbf{D}}
\newcommand{\bE}{\mathbf{E}}

\newcommand{\bF}{\mathbf{F}}

\newcommand{\bh}{\mathbf{h}}

\newcommand{\bK}{\mathbf{K}}

\newcommand{\bm}{\mathbf{m}}
\newcommand{\bP}{\mathbf{P}}

\newcommand{\bu}{\mathbf{u}}

\newcommand{\bv}{\mathbf{v}}

\newcommand{\bx}{\mathbf{x}}
\newcommand{\bX}{\mathbf{X}}
\newcommand{\by}{\mathbf{y}}
\newcommand{\bY}{\mathbf{Y}}
\newcommand{\bV}{\mathbf{V}}
\newcommand{\bW}{\mathbf{W}}
\newcommand{\bz}{\mathbf{z}}

\renewcommand{\epsilon}{\varepsilon}
\renewcommand{\hat}{\widehat}
\renewcommand{\tilde}{\widetilde}

\newcommand{\distn}[1]{\mathcal{#1}}

\newcommand{\Em}{\mathbb E}

\newcommand{\gvn}{\,|\,}
\newcommand{\e}{\text{e}}

\newcommand{\vect}[1]{\boldsymbol #1}

\newcommand{\valpha}{\vect{\alpha}}

\newcommand{\vkappa}{\vect{\kappa}}

\newcommand{\vphi}{\vect{\phi}}

\newcommand{\vepsilon}{\vect{\epsilon}}

\newcommand{\vOmega}{\vect{\Omega}}
\newcommand{\vomega}{\vect{\omega}}
\newcommand{\vSigma}{\vect{\Sigma}}

\newcommand{\vpsi}{\vect{\psi}}

\newcommand{\bigcdot}{\boldsymbol{\cdot}}
\newcommand{\matlab}{\mathrm{M}\mathrm{{\scriptstyle ATLAB}}}

\bibpunct{(}{)}{;}{a}{,}{,}

\begin{document}

\title{Large Order-Invariant Bayesian VARs with \\ Stochastic Volatility\thanks{We would like to thank Chenghan Hou for many helpful comments.}}

\author{Joshua C. C. Chan\\
	{\small Purdue University} \\
	\and Gary Koop \\
	{\small University of Strathclyde} \\
	\and Xuewen Yu\\
	{\small Purdue University}
}

\date{October 2021}

\maketitle

\onehalfspacing

\begin{abstract}

\noindent Many popular specifications for Vector Autoregressions (VARs) with multivariate stochastic volatility are not invariant to the way the variables are ordered due to the use of a Cholesky decomposition for the error covariance matrix. We show that the order invariance problem in existing approaches is likely to become more serious in large VARs. We propose the use of a specification which avoids the use of this Cholesky decomposition. We show that the presence of multivariate stochastic volatility allows for identification of the proposed model and prove that it is invariant to ordering. We develop a Markov Chain Monte Carlo algorithm which allows for Bayesian estimation and prediction. In exercises involving artificial and real macroeconomic data, we demonstrate that the choice of variable ordering can have non-negligible effects on empirical results. In a macroeconomic forecasting exercise involving VARs with $20$ variables we find that our order-invariant approach leads to the best forecasts and that some choices of variable ordering can lead to poor forecasts using a conventional, non-order invariant, approach. 
\bigskip


\bigskip


\end{abstract}

\thispagestyle{empty}


\newpage

\section{Introduction}

Vector Autoregressions (VARs) are one of the most popular models in modern macroeconomics. In recent years, two prominent developments have occured in the Baysian VAR literature. First, beginning with \citet{BGR10} researchers have been working with large VARs involving dozens or even hundreds of dependent variables. Second, there is an increasingly recognition, in most macroeconomic datasets, of the need to allow for time variation in VAR error variances, see, among many others, \citet{clark11}. Both of these developments have led to VAR specifications which are not invariant to the way the variables are ordered in the VAR. This does not matter in the case where a logical ordering of the variables suggests itself (e.g. as in some structural VARs) or when the number of variables in the VAR is very small since the researcher can consider all possible ways of ordering the variables. But with a large number of variables it is unlikely that a logical ordering suggests itself and it is not practical to consider every possible ordering. These considerations motivate the present paper. In it we describe an alternative VAR specification with stochastic volatility (SV) and prove that it is invariant to the ways the variables are ordered. We develop a computationally-efficient Markov Chain Monte Carlo (MCMC) algorithm which is scaleable and allows for Bayesian estimation and forecasting even in high dimensional VARs. We carry out empirical work with artificial and real data which demonstrates the consequences of a lack of order invariance in conventional approaches and that our approach does not suffer from them. 

There are two basic insights that underlie our approach. The first is that it is the use of a Cholesky decomposition of the error covariance matrix $\vSigma_t$ that leads to order dependence. Beginning with influential early VAR-SV papers such as  \citet{CS05} and \citet{primiceri05}, this Cholesky decomposition has been used in numerous papers. The second is that SV can be used to identify structural VARs, see \citet{BB20}. We adopt a VAR-SV specification that avoids the use of the Cholesky decomposition. If we were working with a homoscedastic VAR, our model would be unidentified. But we let the SV achieve identification. In theory we prove order invariance and, in practice, we show that it works well. 

The remainder of this paper is organized as follows. The next section discusses ordering issues in VAR-SVs. The third section introduces our multivariate SV process and proves that it is identified and order invariant. The fourth section adds the VAR component to the model and discusses Bayesian estimation of the resulting order-invariant VAR-SV. The fifth section uses artificial data to illustrate the performance of our model in terms of accuracy and computation time. The sixth section investigates the practical consequences of using the non-order invariant approach of  \citet{CS05} using artificial and real macroeconomic data. Finally, section seven carries out a forecasting exercise which demonstrates that our order-invariant approach forecasts well, but that forecasts produced by the model of \citet{CS05} are sensitive to the way that the variables are ordered and some ordering choices can lead to poor forecast performance. 

\section{Ordering Issues in VAR-SVs}

Ordering issues in VARs relate to the error covariance matrix, not the VAR coefficients themselves. For instance, \citet{CCM19} use a triangular VAR-SV which involves a Cholesky decomposition of the error covariance matrix, $\vSigma_t$. They demonstrate that, conditional on $\vSigma_t$, the draws of the VAR coefficients are invariant to ordering.\footnote{It is worth noting that one major reason for triangularizing the system is to allow for equation by equation estimation of the model. This leads to large computational benefits since there is no need to manipulate the (potentially enormous) posterior covariance matrix of the VAR coefficients in a single block. In particular, \citet{CCM19} show how, relative to full system estimation, triangularisation can reduce computational complexity of Bayesian estimation of large VARs from $O(n^6)$ to $O(n^4)$ where $n$ is VAR dimension. This highlights the importance of development of VAR specifications, such as the one in this paper, which allow for equation by equation estimation. } Such invariance to ordering of the VAR coefficients holds for all the models discussed in the paper. Accordingly, in this section and the next we focus on the multivariate stochastic volatility process. We say a model is invariant to ordering if the order of the variables does not affect the posterior distribution of the error covariance matrix, subject to permutations. In other words, if the ordering of variables $i$ and $j$ in the VAR is changed, then posterior inference on the error covariances and variances associated with variables $i$ and $j$ is unaffected for all $i$ and $j$.

Before discussing multivariate stochastic volatility approaches which are not order invariant, it is worthwhile noting that there are some multivariate SV approaches which are order invariant. Wishart or inverse-Wishart processes (e.g., \citet{PG2006}, \citet{AM09} and \citet{CDLS18}) are by construction invariant to ordering. However, estimation of SV models based on these multivariate processes is computationally intensive as it typically involves drawing from high-dimensional, non-standard distributions. Consequently, these models are generally not applicable to large datasets. In contrast, the common stochastic volatility models considered in \citet{CCM16} and \citet{chan20} are order invariant and can be estimated quickly even for large systems. However, these models are less flexible as the time-varying error covariance matrix depends on a single SV process (e.g., error variances are always proportional to each other). Another order invariant approach is based on the discounted Wishart process (e.g., \citet{Uhlig97}, \citet{WH06}, \citet{PW10} and \citet{Bognanni18}), which admits efficient filtering and smoothing algorithms for estimation. However, this approach appears to be too tightly parameterized for macroeconomic data and it tends to underperform in terms of both point and density forecasts relative to standard SV models such as \citet{CS05} and \citet{primiceri05}; see, for example, \citet{ARRS21} for a forecast comparison exercise.

Another model that is potentially order invariant is the factor stochastic volatility model, and recently these models have been applied to large datasets. However, factor models have some well-known drawbacks such as mis-specification concerns associated with assuming a high dimensional volatility process is well approximated with a low dimensional set of factors. \citet{kastner19} notes that, if one imposes the usual identification restrictions (e.g., factor loadings are triangular), then the model is not order invariant. \citet{kastner19} does not impose those restrictions, arguing that identification of the factor loadings is not necessary for many purposes (e.g. forecasting).  

The multivariate stochastic volatility specification that is perhaps most commonly used in macroeconomics is that of \citet{CS05}.  The SV specification of the reduced-form errors, the $n$-dimensional vector $\bu_t$, in \citet{CS05} can be written as 
\begin{equation}\label{eq:CS0}
	\bu_t = \bB_0^{-1} \vepsilon_t, \quad \vepsilon_t \sim\distn{N}(\mathbf{0},\bD_t),
\end{equation}
where $\bB_0$ is a unit lower triangular matrix with elements $b_{ij}$ and $\bD_t = \text{diag}(\e^{h_{1,t}},\ldots,\e^{h_{n,t}})$ is a diagonal matrix consisting of the log-volatility $\bh_t=(h_{1,t}, \ldots, h_{n,t})'$. By assuming that $\bB_0$ is lower triangular, the implied covariance matrix on $\bu_t$ naturally depends on the order of the elements in $\bu_t$. In particular, one can show that under standard assumptions, the variance of the $i$-th element of $\bu_t$, $u_{i,t}$, increases as $i$ increases. Hence, this ordering issue becomes worse in higher dimensional models. In addition, since 
the stochastic volatility specification in \citet{primiceri05} is based on \citet{CS05} but with $\bB_0$ time-varying, it has a similar ordering issue.

To illustrate this key point, we assume for simplicity that $\vepsilon_t\sim\distn{N}(\mathbf{0},\mathbf{I}_n)$. We further assume that the priors on the lower triangular elements of $\bB_0$ are independent with prior means 0 and variances 1. Since $\bB_0\bu_t = \vepsilon_t$, we can obtain the elements of $\bu_t$ recursively and compute their variances $\text{Var}(u_{i,t}) = \Em u_{i,t}^2$. In particular, we have
\begin{align*}
	u_{1,t} & = \epsilon_{1,t} \\
	u_{2,t} & = \epsilon_{2,t} - b_{2,1}u_{1,t} \\
	u_{3,t} & = \epsilon_{3,t} - b_{3,2}u_{2,t} - b_{3,1}u_{1,t} \\
	       & \vdots \\
	u_{i,t} & = \epsilon_{i,t} - b_{i,i-1}u_{i-1,t} - \cdots - b_{i,1}u_{1,t}.
\end{align*}
Since $\Em(b_{i,j}b_{i,k}) = 0$ for all $j \neq k$, the expectation of any cross-product term in $u_{i,t}^2$ is 0. Hence, we have
\begin{align*}
	\Em u_{1,t}^2 & = \Em \epsilon_{1,t}^2 = 1 \\
	\Em u_{2,t}^2 & = \Em \epsilon_{2,t}^2 + \Em b_{2,1}^2 \Em u_{1,t}^2 = 1 + \Em u_{1,t}^2 = 2  \\
	\Em u_{3,t}^2 & = \Em \epsilon_{3,t}^2 + \Em b_{3,2}^2\Em u_{2,t}^2 + \Em b_{3,1}^2\Em u_{1,t}^2 = 
	1 + \Em u_{2,t}^2 + \Em u_{1,t}^2 = 4 \\
	& \vdots \\
	\Em u_{i,t}^2 & = \Em \epsilon_{i,t}^2 + \Em b_{i,i-1}^2 \Em u_{i-1,t}^2 + \cdots +
	\Em b_{i,1}^2\Em u_{1,t}^2 = 2^{i-1}.
\end{align*}
In other words, the variance of the reduced-form error increases (exponentially) as it is ordered lower in the $n$-tuple, even though the assumptions about $b_{i,j}$ and $\epsilon_{i,t}$ are identical across $i = 1,\ldots, n$. 



\section{An Order-Invariant Stochastic Volatility Model} \label{s:SV}

\subsection{Model Definition}

In this section we extend the stochastic volatility specification in \citet{CS05} with the goal of constructing an order-invariant model. The key reason why the ordering issue in \citet{CS05} occurs is because of the lower triangular assumption for $\bB_0$. A simple solution to this problem is to avoid this assumption and assume that $\bB_0$ is an unrestricted square matrix. To that end, let $\by_t= (y_{1,t},\ldots,y_{n,t})'$ be an $n\times 1$ vector of variables that is observed over the periods $t=1,\ldots, T.$ To fix ideas, consider the following model with zero conditional mean:
\begin{equation}\label{eq:CS}
	\by_t = \bB_0^{-1} \vepsilon_t, \quad \vepsilon_t \sim\distn{N}(\mathbf{0},\bD_t),
\end{equation}
where $\bD_t = \text{diag}(\e^{h_{1,t}},\ldots,\e^{h_{n,t}})$ is diagonal. Here we assume that $\bB_0$ is non-singular, but is otherwise unrestricted. Each of the log-volatilities follows a stationary AR(1) process:
\begin{equation}\label{eq:h}
	h_{i,t} = \phi_i h_{i,t-1} + u_{i,t}^h, \quad u_{i,t}^h\sim\distn{N}(0,\omega_i^2),
\end{equation}
for $t=2,\ldots, T$, where $|\phi_i|<1$ and the initial condition is specified as 
$h_{i,1}\sim\distn{N}(0,\omega_i^2/(1-\phi_i^2))$. Note that the unconditional mean of the AR(1) process is normalized to be zero for identification. 
We call this the Order Invariant SV (OI-SV) model.

\subsection{Identification and Order-Invariance}

The question of identification of our OI-SV model can be dealt with quickly. If there is no SV, it is well-known that $\bB_0$ is not identified. In particular, any orthogonal transformation of $\bB_0$ gives the same likelihood. But with the presence of SV, \citet{BB20} show that $\bB_0$ is identified up to permutations and sign changes:
\begin{proposition}[\citet{BB20}] \label{thm:BB20} \rm Consider the stochastic volatility model given in \eqref{eq:CS}-\eqref{eq:h} with $ \phi_i\neq 0, i=1,\ldots,n.$ Then,  $\bB_0$ is unique up to permutation of its column and multiplication of its columns by $-1$.
\end{proposition}
The assumptions in Proposition \ref{thm:BB20} are stronger than necessary for identification. In fact, \citet{BB20} show that having $n-1$ SV processes is sufficient for identification and derive results relating to partial identification which occurs if there are $n-r$ SV processes  with $r>1$. Intuitively, allowing for time-varying volatility generates non-trivial autocovariances in the squared reduced form errors. These additional moments help identify $\bB_0$.
We refer readers to \citet{LLM10} and \citet{Lewis21} for a more detailed discussion of how identification can be achieved in structural VARs via heteroscedasticity.

Below we outline a few theoretical properties of the stochastic volatility model described in \eqref{eq:CS}. First, we show that the model is order invariant, in the sense that if we permute the order of the dependent variables, the likelihood function implied by this new ordering is the same as that of the original ordering, provided that we permute the parameters accordingly. Stacking $\by = (\by_1',\ldots, \by_T')'$ and $\bh = (\bh_1',\ldots, \bh_T')'$, note that the likelihood implied by~\eqref{eq:CS} is
\[
	p(\by\gvn\bB_0,\bh) = (2\pi)^{-\frac{nT}{2}}|\det \bB_0|^T\prod_{t=1}^T|\det \bD_t|^{-\frac{1}{2}}
	\e^{-\frac{1}{2}\sum_{t=1}^T\by_t'\bB_0'\bD_t^{-1}\bB_0\by_t},
\]
where $|\det \bC |$ denotes the absolute value of the determinant of the square matrix $\bC$. Now, suppose we permute the order of the dependent variables. More precisely, let $\bP$ denote an arbitrary permutation matrix of dimension $n$, and we define $\tilde{\by}_t = \bP\by_t$. By the standard change of variable result, the likelihood of $\tilde{\by} = (\tilde{\by}_1',\ldots, \tilde{\by}_T')'$ is
\begin{align*}
	|\det\bP^{-1}|^{T}& (2\pi)^{-\frac{nT}{2}}|\det \bB_0|^T\prod_{t=1}^T|\det \bD_t|^{-\frac{1}{2}}
	\e^{-\frac{1}{2}\sum_{t=1}^T(\bP^{-1}\tilde{\by}_t)'\bB_0'	\bD_t^{-1}\bB_0\bP^{-1}\tilde{\by}_t} \\
	& = (2\pi)^{-\frac{nT}{2}} |\det \bB_0\bP'|^T\prod_{t=1}^T|\det \bD_t|^{-\frac{1}{2}}\e^{-\frac{1}{2}\sum_{t=1}^T\tilde{\by}_t'\bP\bB_0'\bD_t^{-1}\bB_0\bP'\tilde{\by}_t} \\
	& = (2\pi)^{-\frac{nT}{2}} |\det \bP \bB_0\bP'|^T\prod_{t=1}^T|\det \bP\bD_t\bP'|^{-\frac{1}{2}}	
	\e^{-\frac{1}{2}\sum_{t=1}^T\tilde{\by}_t'\bP\bB_0'\bP'\bP\bD_t^{-1} \bP'\bP\bB_0\bP'\tilde{\by}_t} \\
	& = (2\pi)^{-\frac{nT}{2}} |\det \tilde{\bB}_0|^T\prod_{t=1}^T|\det \tilde{\bD}_t|^{-\frac{1}{2}}
		\e^{-\frac{1}{2}\sum_{t=1}^T\tilde{\by}_t'\tilde{\bB}_0' \tilde{\bD}_t^{-1} \tilde{\bB}_0\tilde{\by}_t} =	p(\tilde{\by}\gvn\tilde{\bB}_0,\tilde{\bh}),
\end{align*}
where $\tilde{\bB}_0 = \bP\bB_0\bP'$, $\tilde{\bD}_t= \bP\bD_t\bP' = \text{diag}(\e^{\tilde{\bh}_t})$ and $\tilde{\bh} = (\tilde{\bh}_1',\ldots, \tilde{\bh}_T')$ with $\tilde{\bh}_t = \bP \bh_t$. Note that we have used the fact that $\bP^{-1} = \bP'$ and $|\det\bP| = 1$, as all permutation matrices are orthogonal matrices. In addition, since $\bP^{-1}\tilde{\by}_t = \by_t$, the first line in the above derivation is equal to $p(\by\gvn\bB_0,\bh)$. Hence, we have shown that $p(\by\gvn\bB_0,\bh) = p(\tilde{\by} \gvn \tilde{\bB}_0,\tilde{\bh})$. We summarize this result in the following proposition. 

\begin{proposition}[Order Invariance] \label{thm:invar} \rm Let $p(\by\gvn\bB_0,\bh)$ denote the likelihood of the stochastic volatility model given in \eqref{eq:CS}. Let $\bP$ be an arbitrary $n\times n$ permutation matrix and define $\tilde{\by}_t = \bP\by_t$ and $\tilde{\bh}_t = \bP \bh_t$. Then, the likelihood with dependent variables $\tilde{\by}_t$ has exactly the same form but with parameters permuted accordingly. More precisely, 
\[
	p(\by\gvn\bB_0,\bh) = p(\tilde{\by} \gvn \tilde{\bB}_0,\tilde{\bh}),
\]
where $\tilde{\bB}_0 = \bP\bB_0\bP'$ and $\tilde{\bh} = (\tilde{\bh}_1',\ldots, \tilde{\bh}_T')$.
\end{proposition}

Next, we derive the unconditional covariance matrix of the data implied by the stochastic volatility model in \eqref{eq:CS}-\eqref{eq:h}. More specifically, given the model parameters $\bB_0$, $\phi_i$ and $\omega^2_i, i=1,\ldots, n,$ suppose we generate the log-volatility processes via \eqref{eq:h}. It is straightforward to compute the implied unconditional covariance matrix of $\by_t$. 

Since the unconditional distribution of $h_{i,t}$ implied by \eqref{eq:h} is $\distn{N}(0,\omega_i^2/(1-\phi_i^2))$, the unconditional distribution of $\e^{h_{i,t}}$ is log-normal with mean 
$\e^{\frac{\omega_i^2}{2(1-\phi_i^2)}}$. Hence, the unconditional mean of $\bD_t$ is 
$\bV_{\bD} = \text{diag}(\e^{\frac{\omega_1^2}{2(1-\phi_1^2)}},\ldots,
\e^{\frac{\omega_n^2}{2(1-\phi_n^2)}})$. It then follows that the unconditional covariance matrix of $\by_t$ can be written as
\[
	\vSigma \equiv (\bB_0'\bV_{\bD}^{-1}\bB_0)^{-1} = (\overline{\bB}_0'\overline{\bB}_0)^{-1},
\]
where $\overline{\bB}_0 = \bV_{\bD}^{-\frac{1}{2}}\bB_0$, i.e., scaling each row of $\bB_0$ by $\e^{-\frac{\omega_i^2}{4(1-\phi_i^2)}}$, $i=1,\ldots, n$. From this expression it is also clear that if we permute the order of the dependent variables via $\tilde{\by}_t = \bP\by_t$, the unconditional covariance matrix of $\tilde{\by}_t$ can be obtained by permuting the rows and columns of $\overline{\bB}_0$ accordingly. More precisely:
\[
	\tilde{\vSigma} \equiv \bP\vSigma\bP' = (\bP\overline{\bB}_0'\bP'\bP\overline{\bB}_0\bP')^{-1} =
	(\tilde{\overline{\bB}}_0'\tilde{\overline{\bB}}_0)^{-1},
\]
where $\tilde{\overline{\bB}}_0 = \bP\overline{\bB}_0\bP'$. Hence, the unconditional variance of any element in $\by_t$ does not depend on its position in the $n$-tuple.

This establishes order-invariance in the likelihood function defined by the OI-SV. Of course, the posterior will also be order invariant (subject to permutations and sign switches) if the prior is. This will hold in any reasonable case. All it requires is that the prior for the parameters in equation $i$ under one ordering becomes the prior for equation $j$ in a different ordering if variable $i$ in the first ordering becomes variable $j$ in the second ordering. 

\section{A VAR with an Order-Invariant SV Specification}

\subsection{Model Definition}
We now add the VAR part to our OI-SV model, leading to the OI-VAR-SV which is given by
\begin{equation} \label{eq:VAR} 
	\by_t = \ba + \bA_1 \by_{t-1} + \cdots + \bA_p \by_{t-p} + \bB_0^{-1}\vepsilon_t, \quad \vepsilon_t \sim\distn{N}(\mathbf{0},\bD_t),
\end{equation}
where $\ba$ is an $n\times 1$ vector of intercepts, $\bA_1, \ldots, \bA_p$ are $n\times n$ matrices of VAR coefficients, $\bB_0$ is non-singular but otherwise unrestricted $n\times n$ matrix, and $\bD_t = \text{diag}(\e^{h_{1,t}},\ldots,\e^{h_{n,t}})$ is diagonal. Finally, each of the log-volatility $h_{i,t}$ follows the stationary AR(1) process as specified in~\eqref{eq:h}.

\subsection{Shrinkage Priors}

We argued previously that ordering issues are likely to be most important in high dimensional models. With large VARs over-parameterization concerns can be substantial and, for this reason, Bayesian methods involving shrinkage priors are commonly used. In this sub-section, we describe a particular VAR prior with attractive properties. But it is worth noting that any of the standard Bayesian VAR priors (e.g. the Minnesota prior) could have been used and the resulting model would still have been order invariant. 

Let $\valpha_i$ denote the VAR coefficients in the $i$-th equation, $i=1,\ldots, n$. We consider the Minnesota-type adaptive hierarchical priors proposed in \citet{chan21}, which have the advantages of both the Minnesota priors (e.g., rich prior beliefs such as cross-variable shrinkage) and modern adaptive hierarchical priors (e.g., heavy-tails and substantial mass around 0). In particular, we construct a Minnesota-type horseshoe prior as follows. For $\alpha_{i,j}$, the $j$-th coefficient in the $i$-th equation, let $\kappa_{i,j} = \kappa_1$ if it is a coefficient on an `own lag' and let $\kappa_{i,j} = \kappa_2$ if it is a coefficient on an `other lag'.
Given the constants $C_{i,j}$ defined below, consider the horseshoe prior on the VAR coefficients (excluding the intercepts) $\alpha_{i,j}, i=1,\ldots,n, j = 2,\ldots, k$ with $k=np+1$:
\begin{align}
	(\alpha_{i,j}\gvn \kappa_1,\kappa_2,\psi_{i,j}) & \sim \distn{N}(m_{i,j},\kappa_{i,j}\psi_{i,j}C_{i,j}), \label{eq:alphaij} \\
	\sqrt{\psi}_{i,j} & \sim \distn{C}^+(0,1), \label{eq:psi} \\
	\sqrt{\kappa}_{1},\sqrt{\kappa}_{2} & \sim \distn{C}^+(0,1), \label{eq:kappa}
\end{align} 
where $\distn{C}^{+}(0,1)$ denotes the standard half-Cauchy distribution. $\kappa_1$ and $\kappa_2$ are the global variance components that are common to, respectively, coefficients of own and other lags, whereas each $\psi_{i,j}$ is a local variance component specific to the coefficient $\alpha_{i,j}$. 
Lastly, the constants $C_{i,j}$ are obtained as in the Minnesota prior, i.e., 
\begin{equation}\label{eq:Cij}
	C_{i,j} = \left\{
	\begin{array}{ll}
			\frac{1}{l^2}, & \text{for the coefficient on the $l$-th lag of variable } i,\\
			\frac{s_i^2}{l^2 s_j^2}, & \text{for the coefficient on the $l$-th lag of variable } j, \; j\neq i,			
	\end{array} \right.
\end{equation}
where $s_r^2$ denotes the sample variance of the residuals from an AR(4) model for the variable~$r, 
r=1,\ldots, n$. Furthermore, for data in growth rates $m_{i,j}$ is set to be 0; for level data, $m_{i,j}$ is set to be zero as well except for the coefficient associated with the first own lag, which is set to be one. 

It is easy to verify that if all the local variances are fixed, i.e., $\psi_{i,j}\equiv 1$, then the Minnesota-type horseshoe prior given in \eqref{eq:alphaij}--\eqref{eq:kappa} reduces to the Minnesota prior. Hence, it can be viewed as an extension of the Minnesota prior by introducing a local variance component such that the marginal prior distribution of $\alpha_{i,j}$ has heavy tails. On the other hand, if $m_{i,j}=0$, $C_{i,j} = 1$ and $\kappa_1 = \kappa_2$, then the Minnesota-type horseshoe prior reduces to the standard horseshoe prior where the coefficients have identical distributions. Therefore, it can also be viewed as an extension of the horseshoe prior \citep{CPS10horseshoe} that incorporates richer prior beliefs on the VAR coefficients, such as cross-variable shrinkage, i.e., shrinking coefficients on own lags differently than other lags.

To facilitate estimation, we follow \citet{MS16} and use the following latent variable representations of the half-Cauchy distributions in \eqref{eq:psi}-\eqref{eq:kappa}:
\begin{align}
	(\psi_{i,j}\gvn z_{\psi_{i,j}}) & \sim \distn{IG}(1/2,1/z_{\psi_{i,j}}), \quad 
	z_{\psi_{i,j}}\sim \distn{IG}(1/2,1),   \label{eq:psi_latent} \\
	(\kappa_{l}\gvn z_{\kappa_{l}}) & \sim \distn{IG}(1/2,1/z_{\kappa_{l}}), \quad 
	z_{\kappa_{l}}\sim \distn{IG}(1/2,1),  \label{eq:kappa_latent} 
\end{align}
for $i=1,\ldots, n, j=2,\ldots, k$ and $l=1,2$ where $\distn{IG}$ denotes the inverse Gamma distribution

Next, we specify the priors on other model parameters. Let $\mathbf{b}_i$ denote the $i$-th row of $\bB_0$ for $i=1,\ldots, n,$ i.e., $\bB_0' = (\mathbf{b}_1,\ldots, \mathbf{b}_n)$. We consider independent priors on $\mathbf{b}_i, i=1,\ldots, n$:
\[	
	\mathbf{b}_i\sim\distn{N}(\mathbf{b}_{0,i},\bV_{\mathbf{b}_i}).
\]
For the parameters in the stochastic volatility equations, we assume the priors for $j=1,\ldots, n$:
\[
	\phi_j\sim \distn{N}(\phi_{0,j},V_{\phi_j})1(|\phi_j|<1),\quad 
	\sigma_{j}^2 \sim \distn{IG}(\nu_{j},S_{j}).
\]

\subsection{MCMC Algorithm}

In this section we develop a posterior sampler which allows for Bayesian estimation of the order-invariant stochastic volatility model with the Minnesota-type horseshoe prior. For later reference, let $\vpsi_i = (\psi_{i,1},\ldots, \psi_{i,k})'$ denote the local variance components associated with $\valpha_i$ and define $\vkappa = (\kappa_1,\kappa_2)'$. Furthermore, let $\bz_{\vpsi_i} = (z_{\psi_{i,1}},\ldots, z_{\psi_{i,k}})'$  denote the latent variables corresponding to $\vpsi_i$ and similarly define $\bz_{\vkappa}$. Next, stack $\by = (\by_1',\ldots, \by_T')'$, $\valpha = (\valpha_1',\ldots, \valpha_n')'$, $\vpsi = (\vpsi_1',\ldots, \vpsi_n')'$, $\bz_{\vpsi} = (\bz_{\vpsi_1}',\ldots, \bz_{\vpsi_n}')'$ and $\bh = (\bh_1',\ldots, \bh_{T}')'$ with $\bh_t = (h_{1,t},\ldots, h_{n,t})'$. Finally, let $\bh_{i,\bigcdot} = (h_{i,1},\ldots, h_{i,T})'$ represent the vector of log-volatility for the $i$-th equation, $i=1,\ldots, n$. Then, posterior draws can be obtained by sampling sequentially from: 
\begin{enumerate}
	\item $p(\bB_0 \gvn \by, \valpha, \bh, \vphi, \vomega^2, \vpsi, \vkappa, \bz_{\vpsi},\bz_{\vkappa})$ = $p(\bB_0 \gvn \by, \valpha, \bh)$; 	
	
	\item $p(\valpha \gvn \by, \bB_0, \bh, \vphi, \vomega^2, \vpsi, \vkappa, \bz_{\vpsi},\bz_{\vkappa})$ = 
	$p(\valpha \gvn \by, \bB_0, \bh, \vpsi, \vkappa)$;
	
	\item $p(\vpsi \gvn \by, \valpha, \bB_0, \bh, \vphi, \vomega^2, \vkappa, \bz_{\vpsi},\bz_{\vkappa})$ = $\prod_{i=1}^n\prod_{j=2}^k p(\psi_{i,j} \gvn \alpha_{i,j}, \vkappa, z_{\psi_{i,j}})$;
	
	\item $p(\vkappa \gvn \by, \valpha, \bB_0, \bh, \vphi, \vomega^2, \vpsi, \bz_{\vpsi},\bz_{\vkappa})$ = $\prod_{l=1}^2 p(\kappa_l \gvn \valpha, \vpsi, \bz_{\kappa_l})$;
	
	\item $p(\bz_{\vpsi} \gvn \by, \valpha, \bB_0, \bh, \vphi, \vomega^2, \vkappa, \vpsi, \bz_{\vkappa})$ = $\prod_{i=1}^n\prod_{j=2}^k p(z_{\psi_{i,j}} \gvn \psi_{i,j})$;
	
	\item $p(\bz_{\vkappa} \gvn \by, \valpha, \bB_0, \bh, \vphi, \vomega^2, \vkappa, \vpsi, \bz_{\vpsi})$ = $\prod_{l=1}^2 p(z_{\kappa_l} \gvn \kappa_{l})$.
	
		\item $p(\bh \gvn \by, \bB_0, \valpha, \vphi,\vomega^2,\vpsi, \vkappa, \bz_{\vpsi},\bz_{\vkappa}) = \prod_{i=1}^{n} p(\bh_{i,\bigcdot} \gvn \by, \bB_0, \valpha, \vphi,\vomega^2)$; 
	
	\item $p(\vomega^2 \gvn \by, \bB_0, \valpha, \bh, \vphi, \vpsi, \vkappa, \bz_{\vpsi},\bz_{\vkappa}) = \prod_{i=1}^{n} p(\omega_i^2 \gvn \bh_{i,\bigcdot}, \phi_i)$;	
	
	\item $p(\vphi \gvn \by, \bB_0, \valpha, \bh, \vomega^2, \vpsi, \vkappa, \bz_{\vpsi},\bz_{\vkappa}) = \prod_{i=1}^{n} p(\phi_i \gvn \bh_{i,\bigcdot}, \sigma^2_i) $;

\end{enumerate}

\textbf{Step 1}. To implement Step 1, we adapt the sampling approach in \citet{WZ03} and \citet{villani09} to the setting with stochastic volatility. More specifically, we aim to draw each row of $\bB_0$ given all other rows. To that end, we first rewrite \eqref{eq:VAR} as:
\begin{equation}\label{eq:VAR_stacked}
	(\bY - \bX\bA)\bB_0' = \bE,
\end{equation}
where $\bY$ is the $T\times n$ matrix of dependent variables, $\bX$ is the $T\times k$ matrix of lagged dependent variables with $k=1+np$, $\bA = (\ba,\bA_1,\ldots, \bA_n)'$ is the $k\times n$ matrix of VAR coefficients and $\bE$ is the $T\times n$ matrix of errors. Then, for $i=1,\ldots, n$, we have
\[
	(\bY - \bX\bA)\mathbf{b}_i = \bE_i, \quad \bE_i \sim\distn{N}(\mathbf{0}, \vOmega_{ \bh_{i,\bigcdot}}),
\]
where $\bE_i$ is the $i$-th column of $\bE$ and $\vOmega_{ \bh_{i,\bigcdot}} = \text{diag}(\e^{h_{i,1}},\ldots, \e^{h_{i,T}}).$ Hence, the full conditional distribution of $\mathbf{b}_i $ is given by
\begin{align}
	p(\mathbf{b}_i\gvn \by, \valpha, \mathbf{b}_{-i}, \bh_{i,\bigcdot}) & \propto 
	|\det \bB_0|^T\e^{-\frac{1}{2}\mathbf{b}_i'(\bY - \bX\bA)'\vOmega_{ \bh_{i,\bigcdot}}^{-1}
	(\bY - \bX\bA)\mathbf{b}_i}\times	\e^{-\frac{1}{2}(\mathbf{b}_i 
	- \mathbf{b}_{0,i})'\bV_{\mathbf{b}_i}^{-1}(\mathbf{b}_i - \mathbf{b}_{0,i})} \nonumber \\
	& \propto  |\det \bB_0|^T \e^{-\frac{1}{2}(\mathbf{b}_i - \hat{\mathbf{b}}_{i})'\bK_{\mathbf{b}_i}
	(\mathbf{b}_i - \hat{\mathbf{b}}_{i})} \label{eq:post_bi}
\end{align}
where 
\[
	\bK_{\mathbf{b}_i} = \bV_{\mathbf{b}_i}^{-1} + (\bY - \bX\bA)'\vOmega_{ \bh_{i,\bigcdot}}^{-1}
	(\bY - \bX\bA), \quad 
	\hat{\mathbf{b}}_{i} = \bK_{\mathbf{b}_i}^{-1}\bV_{\mathbf{b}_i}^{-1}\mathbf{b}_{0,i}.
\]
The above full conditional posterior distribution is non-standard and direct sampling is infeasible. Fortunately, \citet{WZ03} develop an efficient algorithm to sample $\mathbf{b}_i$ in the case when $\hat{\mathbf{b}}_{i}  = \mathbf{0}.$ \citet{villani09} further generalizes this sampling approach for non-zero $\hat{\mathbf{b}}_{i}$. 

The key idea is to show that $\mathbf{b}_{i}$ has the same distribution as a linear transformation of an $n$-vector that consists of one absolute normal and $n-1$ normal random variables. A random variable $Z$ follows the absolute normal distribution $\distn{AN}(\mu,\rho)$ if it has the density function
\[
	f_{\distn{AN}}(z;\mu,\rho) = c|z|^{\frac{1}{\rho}}\e^{-\frac{1}{2\rho}(z-\mu)^2}, \quad 
	z\in\mathbb{R}, \rho\in\mathbb{R}^{+}, \mu\in\mathbb{R},
\]
where $c$ is a normalizing constant. 

To formally state the results, let $\bC_i$ denote the Cholesky factor of $T^{-1}\bK_{\mathbf{b}_i}$ such that $\bK_{\mathbf{b}_i} = T\bC_i \bC_i'$. Let $\bF_{-i}$ represent the matrix $\bF$ with the $i$-th column deleted and $\bF^{\perp} $ be the orthogonal complement of $\bF$. Furthermore, define $\bv_1 = \bC_i^{-1}(\bB_{0,-i}')^{\perp} / ||\bC_i^{-1}(\bB_{0,-i}')^{\perp}||$, where $(\bB_{0,-i}')^{\perp} \equiv ((\bB_{0}')_{-i})^{\perp} $, and let $(\bv_2,\ldots, \bv_n) = \bv_{1}^{\perp}$. Then, we claim that
\[
	(\mathbf{b}_i\gvn \by, \valpha, \mathbf{b}_{-i}, \bh_{i,\bigcdot}) \stackrel{\text{d}}{=}
	(\bC_i')^{-1}\sum_{j=1}^n \xi_j\bv_j,
\]
where $\xi_1\sim \distn{AN}(\hat{\xi}_1,T^{-1})$ and $\xi_j \sim \distn{N}(\hat{\xi}_j,T^{-1}), j=2,\ldots, n$ with $\hat{\xi}_j = \hat{\mathbf{b}}_i'\bC_i\bv_j$. 

To prove the claim, let $\mathbf{b}_i = (\bC_i')^{-1}\sum_{j=1}^n \xi_j\bv_j$. It suffices to show that if we substitute $\mathbf{b}_i$ into \eqref{eq:post_bi}, $\xi_1,\ldots, \xi_n$ are independent with $\xi_1\sim \distn{AN}(\hat{\xi}_1,T^{-1})$ and $\xi_j \sim \distn{N}(\hat{\xi}_j,T^{-1}), j=2,\ldots, n$. To that end, note that by construction, $\bv_1,\ldots, \bv_n$ forms an orthonormal basis of $\mathbb{R}^n$, particularly $\sum_{j=1}^n\bv_j\bv_j' = \mathbf{I}_n$. We first write the quadratic form in  \eqref{eq:post_bi} as:
\begin{align*}
	(\mathbf{b}_i - \hat{\mathbf{b}}_{i})'\bK_{\mathbf{b}_i}(\mathbf{b}_i - \hat{\mathbf{b}}_{i})
	& = T\left(\sum_{j=1}^n\xi_j\bv_j - \bC_i'\hat{\mathbf{b}}_{i}\right)'
	\left(\sum_{j=1}^n\xi_j\bv_j - \bC_i'\hat{\mathbf{b}}_{i}\right)\\
	& = T\left(\sum_{j=1}^n\xi_j^2 - 2\sum_{j=1}^n\xi_j\hat{\mathbf{b}}_{i}'\bC_i\bv_j 
	+ \hat{\mathbf{b}}_{i}'\bC_i\left(\sum_{j=1}^n\bv_j\bv_j'\right) \bC_i\hat{\mathbf{b}}_{i}\right)\\
	& = T\sum_{j=1}^n(\xi_j - \hat{\xi}_j)^2.
\end{align*}
Next, note that by construction, $(\bv_2,\ldots, \bv_n)$ spans the same space as $\bB_{0,-i}'$. Hence, it follows that 
\[
	|\det \bB_0| = |\det \bB_0'|  = 
	|\det (\mathbf{b}_1,\ldots, \mathbf{b}_{i-1}, (\bC_i')^{-1}\sum_{j=1}^n \xi_j\bv_j,
	\mathbf{b}_{i+1},\ldots, \mathbf{b}_{n})| \propto |\xi_1|.
\]
Finally, substituting the quadratic form and the determinant into  \eqref{eq:post_bi}, we obtain
\begin{align*}
	p(\mathbf{b}_i\gvn \by, \valpha, \mathbf{b}_{-i}, \bh_{i,\bigcdot}) 
	\propto &  |\xi_1|^T \e^{-\frac{T}{2}\sum_{j=1}^n(\xi_j - \hat{\xi}_j)^2} \\
	= & |\xi_1|^T \e^{-\frac{T}{2}(\xi_1 - \hat{\xi}_1)^2}\prod_{j=2}^n
		\e^{-\frac{T}{2}(\xi_j - \hat{\xi}_j)^2}.
\end{align*}
In other words, $\xi_1\sim \distn{AN}(\hat{\xi}_1,T^{-1})$ and $\xi_j \sim \distn{N}(\hat{\xi}_j,T^{-1}), j=2,\ldots, n$, and we have proved the claim.

In order to use the above result, we need an efficient way to sample from $\distn{AN}(\hat{\xi}_1,T^{-1})$. This can be done by using the 2-component normal mixture approximation considered in Appendix~C of \citet{villani09}. In addition, the orthogonal complement of $\bv_1$, namely, $\bv_{1}^{\perp}$, can be obtained using the singular value decomposition.\footnote{In \textsc{MATLAB}, the orthogonal complement of $\bv_1$ can be obtained using \texttt{null}$(\bv_1')$.} In the sampler we fix the sign of the $i$-th element of $\mathbf{b}_i$ to be positive (so that the diagonal elements 
of $\bB_0$ are positive). This is done by multiplying the draw $\mathbf{b}_i$ by $-1$ if its $i$-th element is negative. 

\textbf{Step 2}. We sample the reduced-form VAR coefficients row by row along the lines of 
\citet{CCM19}. In particular, we extend the triangular algorithm in \citet{CCCM21} that is designed for a lower triangular impact matrix $\bB_0$ to the case where $\bB_0$ is a full matrix. To that end, define $\bA_{i=0}$ to be a 
$k\times n$ matrix that has exactly the same elements as $\bA$ except for the $i$-th column, which is set to be zero, i.e., $\bA_{i=0} = (\valpha_1,\ldots, \valpha_{i-1}, \mathbf{0}, \valpha_{i+1},\ldots, \valpha_n).$ Then, we can rewrite~\eqref{eq:VAR} as
\[
	\bB_0(\by_t - \bA_{i=0}'\bx_t) = (\bB_{0,i}\otimes\bx_t')\valpha_i + \vepsilon_t, \quad \vepsilon_t\sim\distn{N}(\mathbf{0},\bD_t),
\]
where $\bx_t = (1,\by_{t-1}',\ldots, \by_{t-p}')'$ and $\bB_{0,i}$ is the $i$-th column of $\bB_0$. Let 
$\bz^i_t = \bB_0(\by_t - \bA_{i=0}'\bx_t)$ and stack $\bz^i = (\bz^{i'}_1,\ldots, \bz^{i'}_T)'$, we have
\[
	\bz^i = \bW^i \valpha_i + \vepsilon, \quad \vepsilon\sim\distn{N}(\mathbf{0},\bD),
\]
where $\bW^i = \bX\otimes\bB_{0,i}$ and $\bD = \text{diag}(\bD_1,\ldots, \bD_T)$.

Next, the Minnesota-type horseshoe prior on $\valpha_i$ is conditionally Gaussian given the local variance component $\vpsi_i$. More specifically, we can rewrite the conditional prior on $\valpha_i$ in~\eqref{eq:alphaij} as:
\[
	(\valpha_i\gvn\vkappa,\vpsi_i)\sim \distn{N}(\bm_i,\bV_{\valpha_i}),
\]
where $\bm_i = (m_{i,1},\ldots, m_{i,k})'$ and $\bV_{\valpha_i} = \text{diag}(C_{i,1},\kappa_{i,2}\psi_{i,2}C_{i,2}\ldots, \kappa_{i,k}\psi_{i,k}C_{i,k})$ (the prior on the intercept is Gaussian). Then, by standard linear regression results, we have
\[
	(\valpha_i \gvn \by, \bB_0, \valpha_{-i}, \bh,  \vpsi_i, \vkappa) \sim \distn{N}(\hat{\valpha}_i,\bK_{\valpha_i}^{-1}),
\]
where $\valpha_{-i}= (\valpha_{1}',\ldots, \valpha_{i-1}', \valpha_{i+1}',\ldots, \valpha_{n}')'$,
\[
	\bK_{\valpha_i} = \bV_{\valpha_i}^{-1} + \bW^{i'}\bD^{-1}\bW^i, \quad 
	\hat{\valpha}_i = \bK_{\valpha_i}^{-1}\left(\bV_{\valpha_i}^{-1}\bm_i	+ \bW^{i'}\bD^{-1}\bz^i\right).
\]
The computational complexity of this step is the same, i.e., $\mathcal{O}(n^4)$, as in the triangular case considered in \citet{CCCM21}.

\textbf{Step 3}. First, note that the elements of $\vpsi_i$ are conditionally independent and we can sample them one by one without loss of efficiency. Next, combining \eqref{eq:alphaij} and \eqref{eq:psi_latent}, we obtain
\begin{align*}
	p(\psi_{i,j} \gvn \alpha_{i,j}, \vkappa, z_{\psi_{i,j}})
		& \propto \psi_{i,j}^{-\frac{1}{2}}\e^{-\frac{1}{2\kappa_{i,j}C_{i,j}\psi_{i,j}}(\alpha_{i,j}-m_{i,j})^2}
		\times \psi_{i,j}^{-\frac{3}{2}}\e^{-\frac{1}{\psi_{i,j}z_{\psi_{i,j}}}} \\
	 & = \psi_{i,j}^{-2}\e^{-\frac{1}{\psi_{i,j}}\left(z_{\psi_{i,j}}^{-1} + \frac{(\alpha_{i,j}-m_{i,j})^2}{2\kappa_{i,j}C_{i,j}}\right)},
\end{align*}
which is the kernel of the following inverse-gamma distribution:
\[
	(\psi_{i,j} \gvn \alpha_{i,j}, \vkappa, z_{\psi_{i,j}})\sim \distn{IG}
	\left(1, z_{\psi_{i,j}}^{-1} + \frac{(\alpha_{i,j}-m_{i,j})^2}{2\kappa_{i,j}C_{i,j}}\right).
\]

\textbf{Step 4}. Note that $\kappa_1$ and $\kappa_2$ only appear in their priors in \eqref{eq:kappa_latent} and in \eqref{eq:alphaij} --- recall $\kappa_{i,j} = \kappa_1$ for coefficients on own lags and $\kappa_{i,j} = \kappa_2$ for coefficients on other lags.  To sample $\kappa_1$ and $\kappa_2$, first define the index set $S_{\kappa_1}$ that collects all the indexes $(i,j)$ such that $\alpha_{i,j}$ is a coefficient associated with an own lag. That is, 
$S_{\kappa_1} = \{(i,j): \alpha_{i,j} \text{ is a coefficient associated with an own lag}\}$. Similarly, define $S_{\kappa_2}$ as the set that collects all the indexes $(i,j)$ such that $\alpha_{i,j}$ is a coefficient associated with a lag of other variables. It is easy to check that the numbers of elements in $S_{\kappa_1}$  and $S_{\kappa_2}$ are respectively $np$ and $(n-1)np$. Then, we have
\begin{align*}
	p(\kappa_1 \gvn \valpha, \vpsi, z_{\kappa_1}) & \propto \prod_{(i,j)\in S_{\kappa_1}} \kappa_1^{-\frac{1}{2}}
	\e^{-\frac{1}{2\kappa_{1}C_{i,j}\psi_{i,j}}(\alpha_{i,j}-m_{i,j})^2}
	\times \kappa_1^{-\frac{3}{2}}\e^{-\frac{1}{\kappa_1 z_{\kappa_1}}} \\
	 & = \kappa_1^{-\left(\frac{np+1}{2}+1\right)} \e^{-\frac{1}{\kappa_2}\left(z_{\kappa_1}^{-1} +
		\sum_{(i,j)\in S_{\kappa_1}}\frac{(\alpha_{i,j}-m_{i,j})^2}{2\psi_{i,j}C_{i,j}}\right)},
\end{align*}
which is the kernel of the $\distn{IG}\left(\frac{np+1}{2},z_{\kappa_1}^{-1} + \sum_{(i,j)\in S_{\kappa_1}}\frac{(\alpha_{i,j}-m_{i,j})^2}{2\psi_{i,j}C_{i,j}}\right)$ distribution. Similarly, we have
\[
	(\kappa_2 \gvn \valpha, \vpsi, z_{\kappa_2}) \sim \distn{IG}\left(\frac{(n-1)np+1}{2}, z_{\kappa_2}^{-1} +
	\sum_{(i,j)\in S_{\kappa_2}} \frac{(\alpha_{i,j}-m_{i,j})^2}{2\psi_{i,j}C_{i,j}}\right).
\]

\textbf{Steps 5-6}. It is straightforward to sample the latent variables $\bz_{\vpsi}$ and $\bz_{\vkappa}$. In particular, it follows from \eqref{eq:psi_latent} that $z_{\psi_{i,j}}\sim\distn{IG}(1,1+\psi_{i,j}^{-1}), i=1,\ldots, n, j=2,\ldots, k$. Similarly, from \eqref{eq:kappa_latent} we have $z_{\kappa_l}\sim\distn{IG}(1,1+\kappa_l^{-1}), l=1,2$. 

\textbf{Step 7}. The log-volatility vector for each equation, $\bh_{i,\bigcdot}, i=1,\ldots, n,$ can be sampled using the auxiliary mixture sampler of \citet{KSC98} once we obtain the orthogonalized errors. More specifically, we first compute $\bE$ using \eqref{eq:VAR_stacked}. Then, we transform the $i$-th column of $\bE$ via $\by_i^* = (\log E_{1,i}^2,\ldots, \log E_{T,i}^2)'$. Finally we implement the auxiliary mixture sampler in conjunction with the precision sampler of \citet{CJ09} to sample $\bh_{i,\bigcdot}$ using $\by_i^*$ as data.

\textbf{Steps 8-9}. These two steps are standard; see, e.g., \citet{CH14}.

\section{Experiments with Artificial Data}\label{s:sim}

\begin{figure}[H]
	\centering	
	\includegraphics[width=1\textwidth]{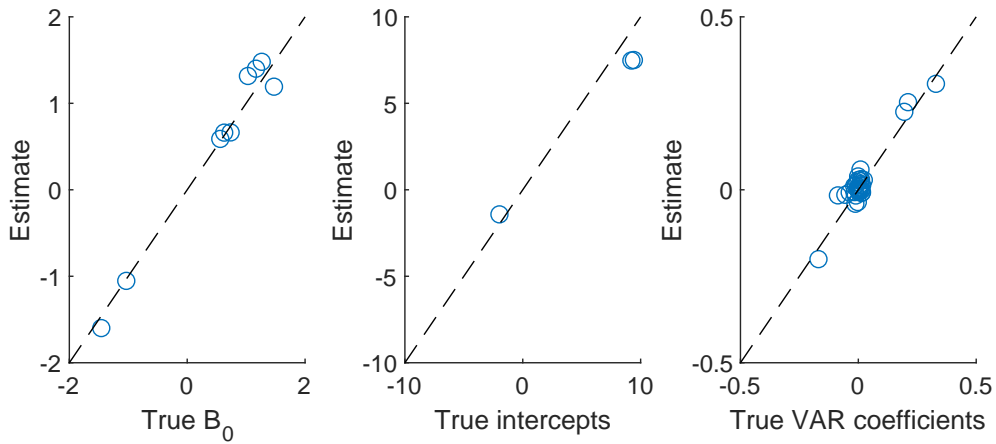}  
	\caption{Scatter plot of the posterior means of $\bB_0$ (left panel), intercepts (middle panel) and VAR coefficients (right panel) against true values from a DGP with stochastic volatility.}
	\label{fig:MC_coef}
\end{figure}

\begin{figure}[H]
	\centering	
	\includegraphics[width=.8\textwidth]{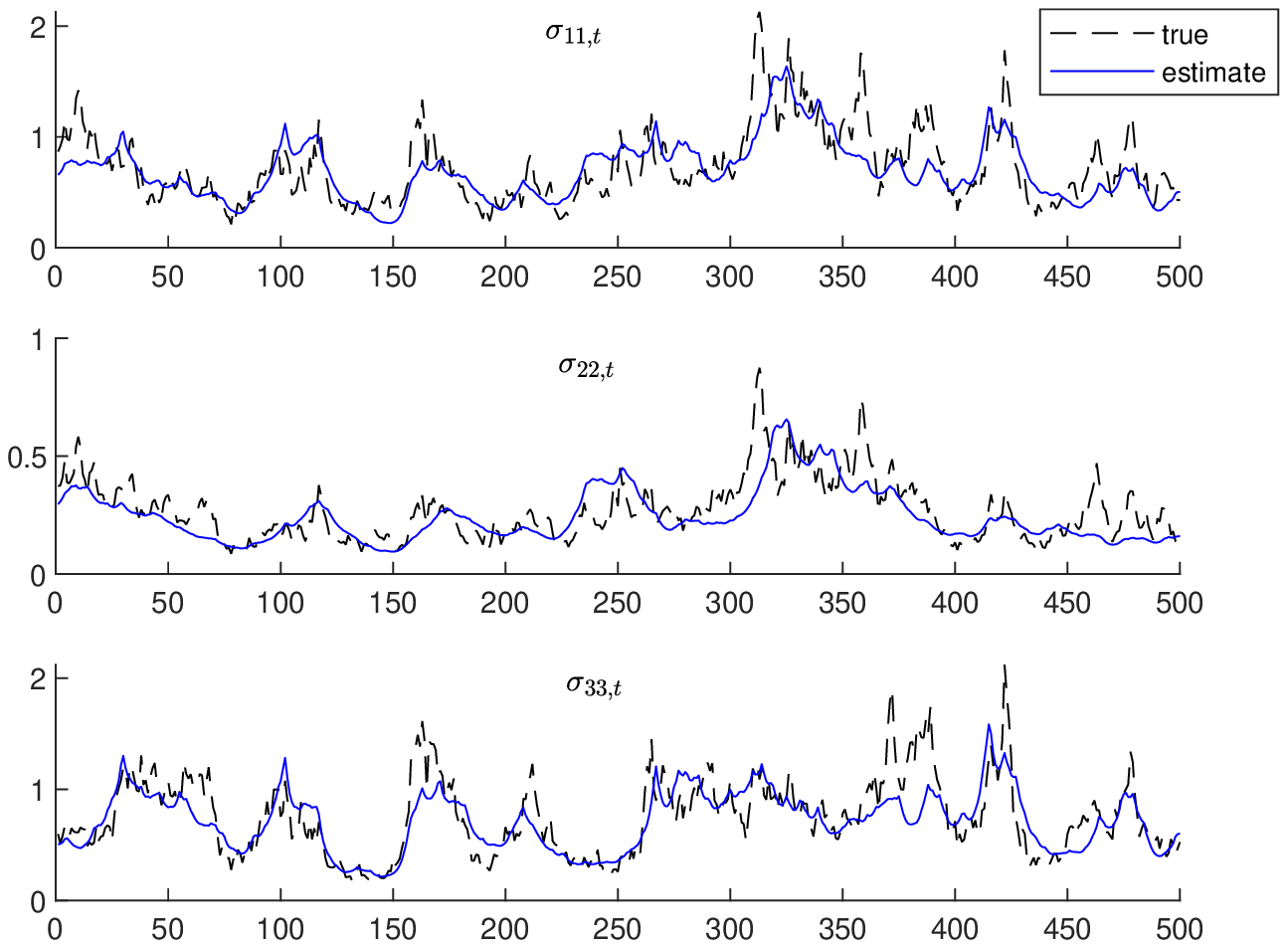}  
		\caption{Estimated time-varying reduced-form variances, where $\sigma_{ii,t}$ denote the $i$-th diagonal element of $\vSigma_t = \bB_0^{-1}\bD_t(\bB_0^{-1})'$, from a DGP with stochastic volatility.}
	\label{fig:MC_Sig1}
\end{figure}

\begin{figure}[H]
	\centering	
	\includegraphics[width=.8\textwidth]{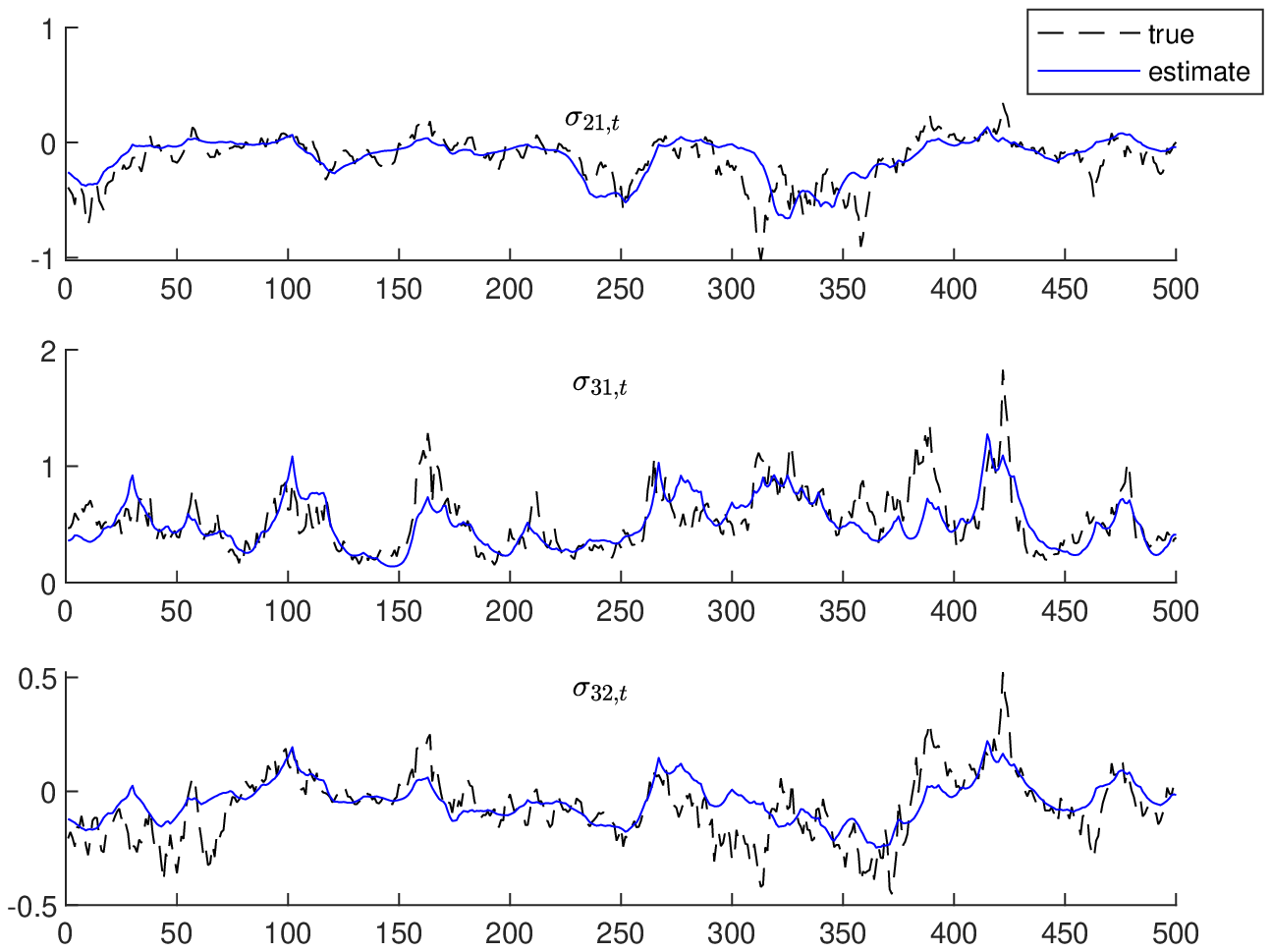}  
		\caption{Estimated time-varying reduced-form covariances, where $\sigma_{ij,t}$ denote the $(i,j)$ element of $\vSigma_t = \bB_0^{-1}\bD_t(\bB_0^{-1})'$, from a DGP with stochastic volatility.}
	\label{fig:MC_Sig2}
\end{figure}

In this section we first present results on two artificial data experiments to illustrate how the new model performs under DGPs with and without stochastic volatility. In the first experiment, we generate a dataset from the VAR in \eqref{eq:VAR} with $n=3$, $T=500$, $p=4$ and the stochastic volatility processes as specified in \eqref{eq:h}. We then estimate the model using the posterior sampler outlined in the previous section. The first dataset is generated as follows. First, the intercepts are drawn independently from $\distn{U}(-10, 10)$, the uniform distribution on the interval $(-10,10)$. For the VAR coefficients, the diagonal elements of the first VAR coefficient matrix are iid $\distn{U}(0,0.5)$ and the off-diagonal elements are from $\distn{U}(-0.2,0.2)$; all other elements of the $j$-th ($j > 1$) VAR coefficient matrices are iid $\distn{N}(0,0.1^2/j^2).$ For the impact matrix $\bB_0$, the diagonal elements are iid $\distn{U}(0.5,2)$, whereas the off-diagonal elements are iid $\distn{N}(0,1)$. Finally, for the SV processes, we set $\phi_i = 0.95$ and $\omega_i^2 = 0.05, i=1,\ldots, n.$ The results of the artificial data experiments are reported in Figures~\ref{fig:MC_coef}-\ref{fig:MC_Sig2}.

Overall, all the estimates track the true values closely. In particular, the new model is able to recover the time-varying reduced-form covariance matrices. In addition, it is also evident that $\bB_0$ can be estimated accurately. 

\begin{figure}[H]
	\centering	
	\includegraphics[width=1\textwidth]{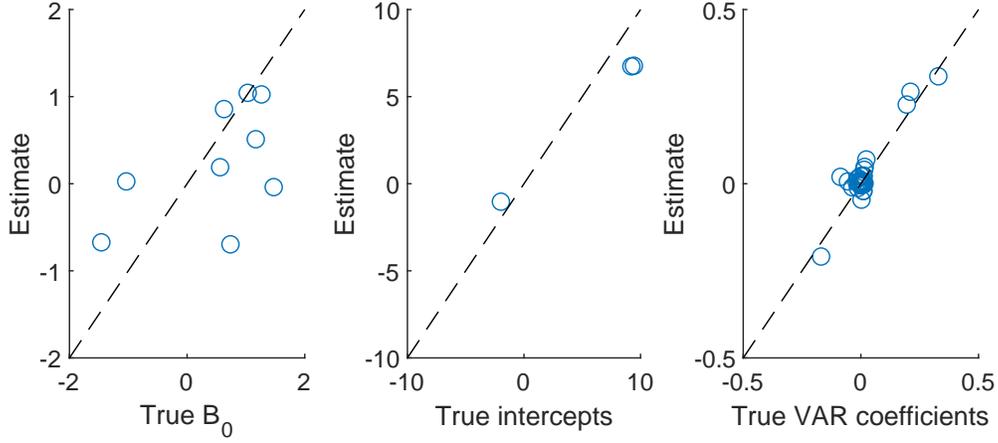}  
	\caption{Scatter plot of the posterior means of $\bB_0$ (left panel), intercepts (middle panel) and VAR coefficients (right panel) against true values from a DGP without stochastic volatility.}
	\label{fig:MC2_coef}
\end{figure}

\begin{figure}[H]
	\centering	
	\includegraphics[width=.8\textwidth]{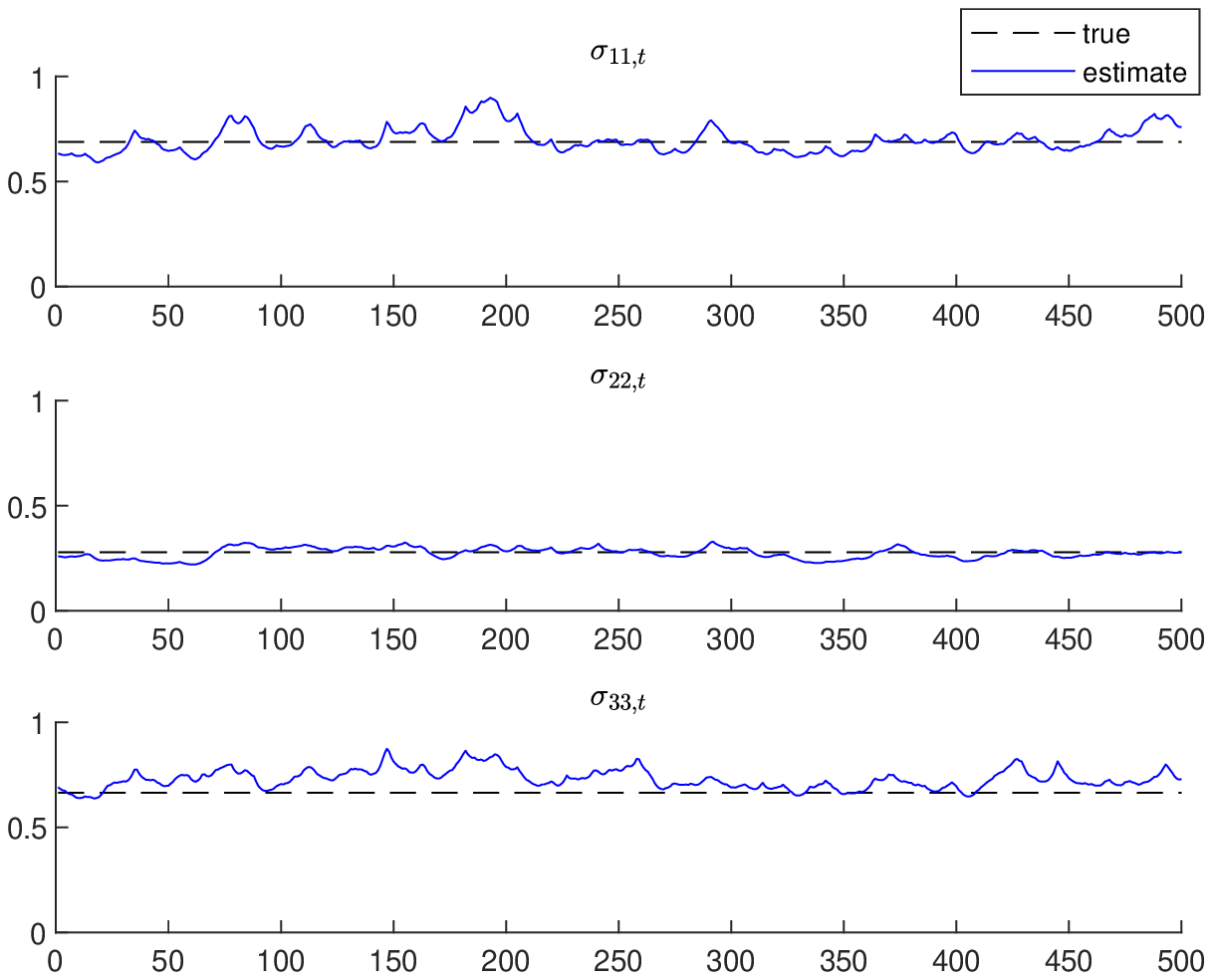}  
		\caption{Estimated time-varying reduced-form variances, where $\sigma_{ii,t}$ denote the $i$-th diagonal element of $\vSigma_t = \bB_0^{-1}\bD_t(\bB_0^{-1})'$, from a DGP without stochastic volatility}
	\label{fig:MC2_Sig1}
\end{figure}

\begin{figure}[H]
	\centering	
	\includegraphics[width=.8\textwidth]{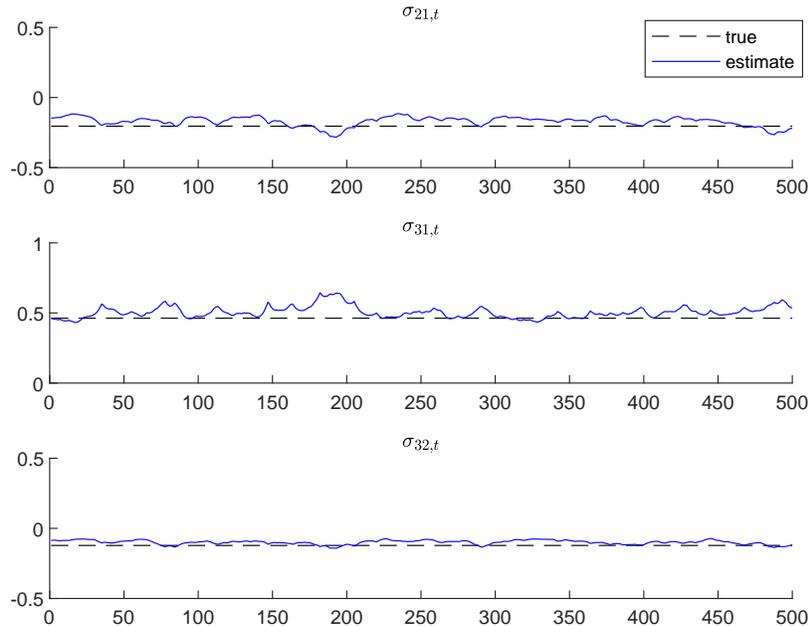}  
		\caption{Estimated time-varying reduced-form covariances, where $\sigma_{ij,t}$ denote the $(i,j)$ element of $\vSigma_t = \bB_0^{-1}\bD_t(\bB_0^{-1})'$, from a DGP without stochastic volatility.}
	\label{fig:MC2_Sig2}
\end{figure}

In the second experiment, we generate data from the same VAR but with the stochastic volatility component turned off so that the errors are homoscedastic (in particular, we set $h_{i,t} = 0$). In other words, we have generated data from a homoscedastic, unidentified model.  But we are estimating it using the (mis-specified) heteroscedastic model with the posterior sampler outlined in the previous sections. The results are reported in Figures~\ref{fig:MC2_coef}-~\ref{fig:MC2_Sig2}. 

When the DGP does not have stochastic volatility, elements of the impact matrix $\bB_0$ are much harder to pin down (as expected since these parameters are not identified). But it is interesting to note that the estimates of the time-varying variances are still able to track the true values fairly closely (as these reduced-form error variances are identified). The VAR coefficients, as well, are well-estimated.

Finally, we document the runtimes of estimating OI-VAR-SV models of different dimensions to assess how well the posterior sampler scales to higher dimensions. More specifically, we report in Table~\ref{tab:times} the computation times (in minutes) to obtain 10,000 posterior draws from the proposed OI-VAR-SV model of dimensions $n= 10, 20, 50$ and sample sizes $T=300, 800$. The posterior sampler is implemented in $\matlab$ on a desktop with an Intel Core i7-7700 @3.60 GHz processor and 64 GB memory. It is evident from the table that for typical applications with 15-30 variables, the OI-VAR-SV model can be estimated quickly. More generally, its estimation time is comparable to that of the triangular model in \citet{CCM19}.

\begin{table}[H]
\centering
\caption{The computation times (in minutes) to obtain 10,000 posterior draws from the proposed order invariant VAR-SV model with $n$ variables and $T$ observations. All VARs have $p = 4$ lags.}
\label{tab:times}
\begin{tabular}{cccccc}
\hline \hline
 \multicolumn{3}{c}{$T = 300$} & \multicolumn{3}{c}{$T = 800$} \\
     $n = 10$  & $n=20$  & $n=50$  & $n = 10$  & $n=20$  & $n=50$  \\ \hline
     2.5  & 17.5 & 233 & 6.8 & 39.7 & 630 \\ \hline \hline
\end{tabular}
\end{table}

\section{Sensitivity to Ordering in the VAR-SV of \citet{CS05}}

In the preceding sections, we have developed Bayesian methods for order-invariant inference in VARs with SV. In this section, we illustrate the importance of this by showing the degree to which results are sensitive to ordering in the VAR-SV of \citet{CS05}, hereafter CS-VAR-SV, which is one of the most popular VAR-SV specifications in empirical macroeconomics. We do this using both simulated and macroeconomic data. To ensure comparability, we use the same prior for the CS-VAR-SV as for our OI-VAR-SV.  
The only difference is the prior on $\bB_0$. For the CS-VAR-SV, $\bB_0$ is unit lower triangular, and we assume the lower triangular elements have prior distribution $\distn{N}(0,1)$. For OI-VAR-SV, we assume the same $\distn{N}(0,1)$ prior for the off-diagonal elements of $\bB_0$, whereas the diagonal elements are iid $\distn{N}(1,1)$. MCMC estimation of the CS-VAR-SV is carried out using the algorithm of \citet{CCCM21}.

\subsection{A Simulation Experiment}

As discussed in \citet{primiceri05}, the Cholesky structure of the model in \citet{CS05} is unable to accommodate certain co-volatility patterns. To illustrate this point, we simulate $T=500$ observations from the model in \eqref{eq:h} and \eqref{eq:VAR} with $p=4$ and $n=3$. To investigate the consequences of the lower triangular restriction we choose a non-triangular data generating process (DGP). We emphasise that the presence of SV in the DGP means that the model is identified. We set 
\[
	\bB_0 = \begin{pmatrix} 1 & -0.8 & -0.8 \\ 0.8 & 1 & -0.8 \\ 0.8 & 0.8 & 1 \end{pmatrix}.
\]

\begin{figure}[H]
	\centering	
	\includegraphics[width=.8\textwidth]{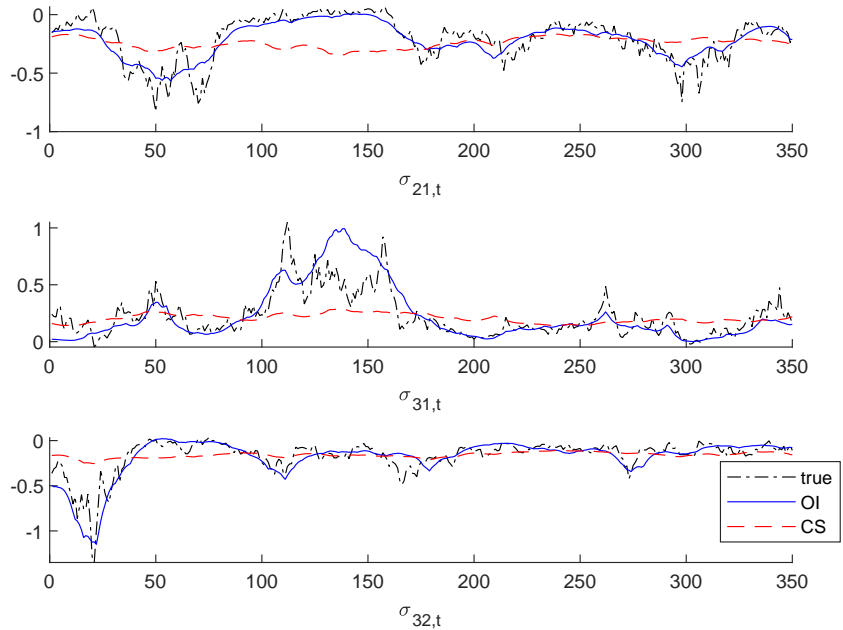}  
	\caption{Estimated time-varying reduced-form covariances from the OI-VAR-SV and CS-VAR-SV models, where $\sigma_{ij,t}$ denote the $(i,j)$ element of $\vSigma_t = \bB_0^{-1}\bD_t(\bB_0^{-1})'.$}
	\label{fig:sim1}
\end{figure}

The VAR coefficients and volatilities are simulated in exactly the same way as described in Section~\ref{s:sim}.
Figure~\ref{fig:sim1} reports the estimated time-varying reduced-form error covariance matrices produced by the CS-VAR-SV and OI-VAR-SV models. 

It is clear from the figure that the estimates from the CS-VAR-SV model tend to be flat and do not capture the time-variation in the error covariances well.	 This is perhaps not surprising as the $\bB_0$ is restricted to be lower triangular in the CS-VAR-SV. However, the OI-VAR-SV does track the true $\sigma_{ij,t}$ quite well. Since error covariances play an important role in features of interest such as impulse responses, this illustrates the potential negative consequences of working with triangularized systems. 

\subsection{Ordering Issues in a 20-Variable VAR}

Next we provide a simple illustration of how the choice of ordering can influence empirical results in large VARs using a dataset of $20$ popular US monthly variables obtained from the FRED-MD database \citep{MN16}. The sample period is from 1959:03 to 2019:12. These variables, along with their transformations are listed in Appendix A. Four core variables, namely, industrial production, the unemployment rate, PCE inflation and the Federal funds rate, are ranked as the first to the fourth variables, respectively. The remaining 16 variables are ordered as in \citet{CCM19}. We estimate the OI-VAR-SV and CS-VAR-SV models with the variables in this order (we call these models OI-VAR-SV-1 and CS-VAR-SV-1). Then we reverse the order of the variables and re-estimate them (these models are OI-VAR-SV-2 and CS-VAR-SV-2).

\begin{figure}[H]
	\centering	
	\includegraphics[width=.7\textwidth]{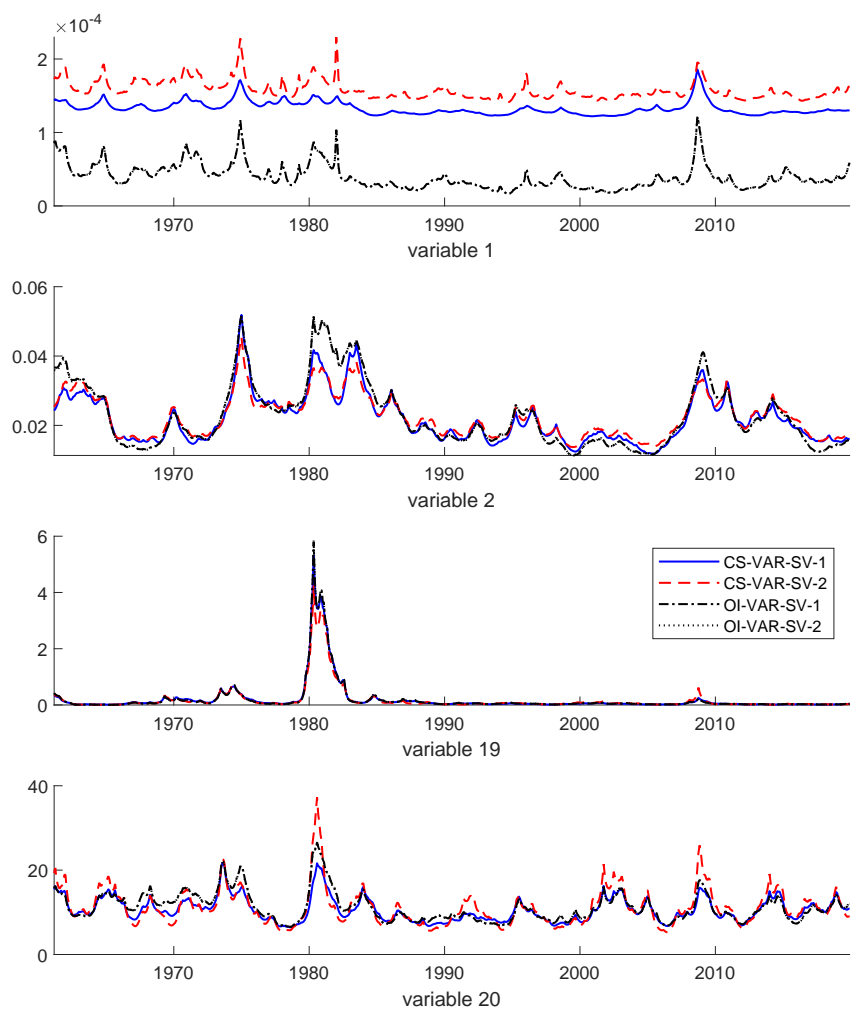}  
	\caption{Estimates of time-varying reduced-form error variances from CS-VAR-SV and OI-VAR-SV under two different  variable orderings}
	\label{fig:sim2}
\end{figure}

\begin{figure}[H]
	\centering	
	\includegraphics[width=.7\textwidth]{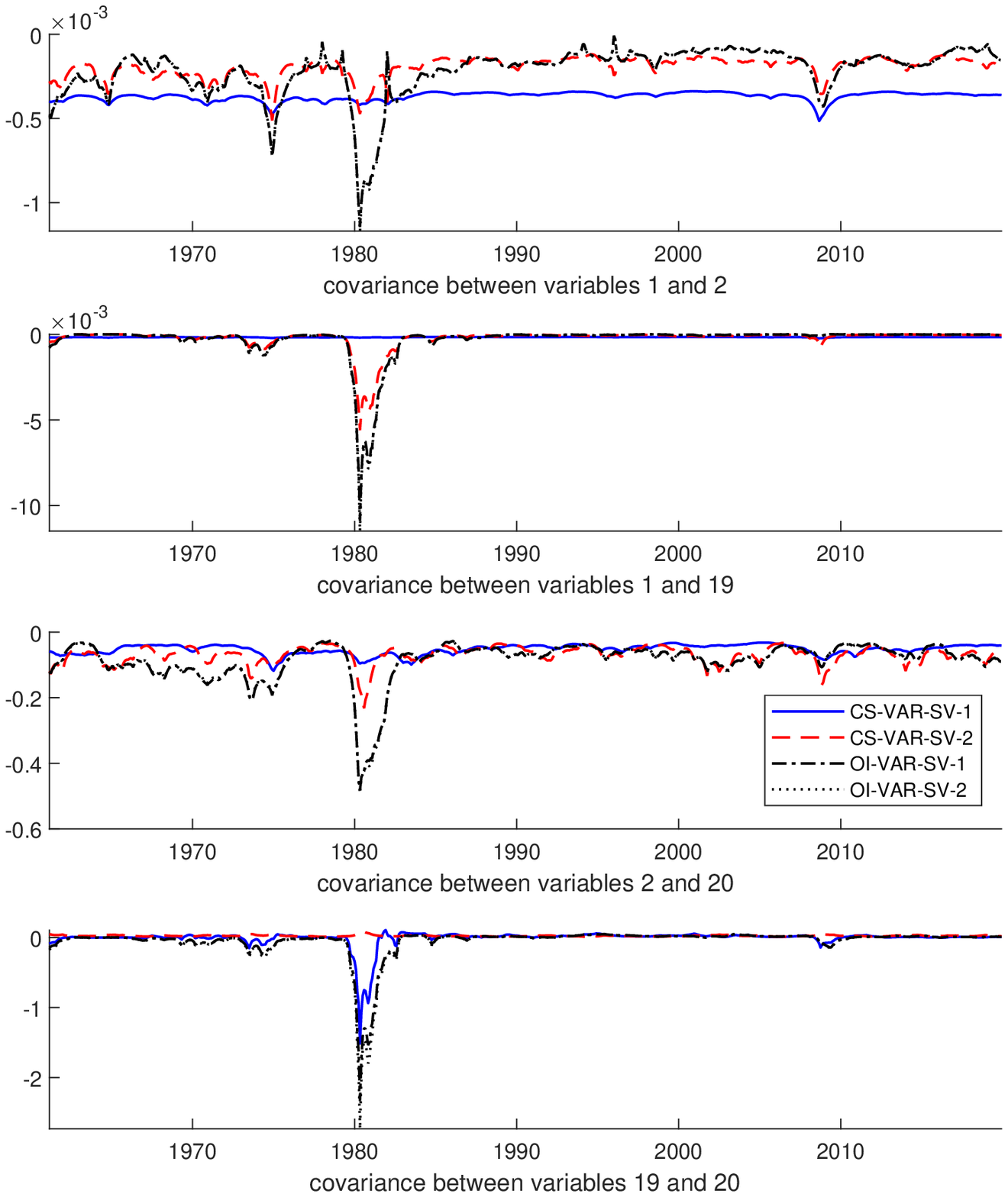}  
	\caption{Estimates of time-varying reduced-form error covariances from from CS-VAR-SV and OI-VAR-SV under two different  variable orderings}
	\label{fig:sim3}
\end{figure}

Estimates (posterior means) of reduced-form variances and covariances of selected variables are reported in Figure~\ref{fig:sim2} and Figure~\ref{fig:sim3}. That is, the first figure reports selected diagonal elements of $\vSigma_t = \bB_0^{-1}\bD_t(\bB_0^{-1})'$ under two different variable orderings, whereas the second figure reports selected off-diagonal elements of $\vSigma_t$. 

There are three points to note about these figures. First, as expected, estimates produced by the OI-VAR-SV model under the two different variable orderings are identical (up to MCMC approximation error). Second, estimates produced by the CS-VAR-SV model under the two different variable orderings are often similar to one another, but occasionally differ, particularly in periods of high volatility. Third, estimates from the CS-VAR-SV models are often similar to the OI-VAR-SV models, but sometimes are substantially different, particularly in the error covariances. 

\begin{figure}[H]
	\centering	
	\includegraphics[width=1\textwidth]{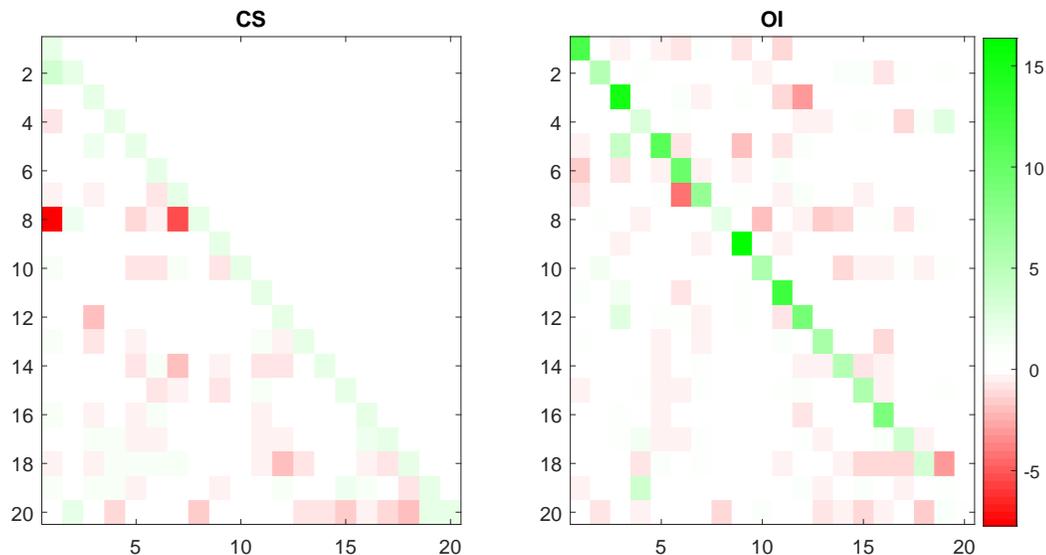}  
	\caption{Posterior means of $\bB_0$ using CS-VAR-SV-1 (left panel) and OI-VAR-SV-1 (right panel)}
	\label{fig1-hm}
\end{figure}

We next present results relating to $\bB_0$ which highlight the differences between our OI-VAR-SV-1 and the CS-VAR-SV-1.\footnote{Note that we are comparing the two models under the first ordering of the variables. The comparable figure using the second ordering reveals similar patterns.}  Figure~\ref{fig1-hm} reports the posterior means of $\bB_0$ of these two models. Note that the two panels of the figure are very different.
Part of this difference is due to the fact that, under OI-VAR-SV, the SV processes are assumed to have zero unconditional means and the diagonal elements of $\bB_0$ are unrestricted. Hence, the diagonal elements of $\bB_0$ play a key role in adjusting the scale of the error variances. In contrast the diagonal elements of $\bB_0$ are restricted to be one for the CS-VAR-SV model. But the key difference lies in the upper-triangular elements of $\bB_0$. These are restricted to be zero in the CS-VAR-SV. But several of these upper triangular elements are estimated to be large in magnitude using our OI-VAR-SV model. The data strongly supports an unrestricted impact matrix $\bB_0$. 

Next, we report the posterior means of the shrinkage hyperparameters $\kappa_1$ and $\kappa_2$ under the two models with two different orderings of the variables. As is clear from the table, the estimates of $\kappa_1$ and $\kappa_2$ can differ substantially when using the CS-VAR-SV model, depending on how the variables are ordered. For example, the estimate of $\kappa_2$ increases by 84\% when one reverses the order of the variables. Of course, under the proposed order-invariant model the estimates are the same (up to MCMC approximation error).

\begin{table}[H]
\centering
\caption{Posterior means of $\kappa_1$ and $\kappa_2$ from the proposed order-invariant stochastic volatility model under two different orderings (OI-VAR-SV-1 and OI-VAR-SV-2), as well as those from the 
model of \citet{CS05} (CS-VAR-SV-1 and CS-VAR-SV-2). }
\begin{tabular}{l|cc|cc}
\hline \hline
          & CS-VAR-SV-1     & CS-VAR-SV-2 & OI-VAR-SV-1 & OI-VAR-SV-2 \\ \hline
$\kappa_1$ & 0.0177  & 0.0170  & 0.0356   & 0.0358   \\
$\kappa_2$ & $2.34\times 10^{-6}$ & $4.30 \times 10^{-6}$ & $2.36\times 10^{-6}$ & $2.25\times 10^{-6}$
\\ \hline     
\end{tabular}
\end{table}

\section{A Forecasting Exercise}

Since VARs are commonly used for forecasting, it is interesting to investigate how the sensitivity to ordering of the CS-VAR-SV models affects forecasts and whether working with our order-invariant specification can lead to forecast improvements. Accordingly we carry out a forecasting exercise using the monthly macroeconomic dataset of sub-section 6.2 for the four models (OI-VAR-SV-1, OI-VAR-SV-2, CS-VAR-SV-1 and CS-VAR-SV-2). Forecast performance of the models is evaluated from 1970:03 till the end of the sample. Root mean squared forecast errors (RMSFEs) are used to evaluate the quality of point forecasts and averages of log predictive likelihoods (ALPLs) are used to evaluate the quality of density forecasts. We focus on four core variables: industrial production, the unemployment rate, PCE inflation and the Federal funds rate. Results for three different forecast horizons ($h=1,6,12$) are presented in Table~\ref{tab:forecasts}. We use the iterated method of forecasting to compute longer horizon forecasts.

Note first that, as expected, OI-VAR-SV-1 and OI-VAR-SV-2 are producing the same forecasts (up to MCMC approximation error). CS-VAR-SV-1 and CS-VAR-SV-2 are often producing forecasts which are substantially different both from one another and from those produced by the order-invariant approaches. These differences are not that noticeable in the RMSFEs, but are very noticeable in the ALPLs. This suggests ordering issues are more important for density forecasts than for point forecasts. This is not surprising since ordering issues are important for the multivariate stochastic volatility process (which plays an important role in determining predictive variances) but not for the VAR coefficients (which are the key determinants of predictive means). These findings are similar to those found in smaller VAR-SVs by \citet{ARRS21}.  

A careful examination of the ALPL results indicates that the best forecasting model is almost always the OI-VAR-SV and in most cases the forecast improvements are statistically significant relative to the CS-VAR-SV-1 benchmark. A comparison of CS-VAR-SV-1 to CS-VAR-SV-2 yields no simple conclusion. Remember that CS-VAR-SV-1 uses a similar ordering of \citet{CCM19}. For most variables and forecast horizons, this ordering is leading to higher ALPLs than the reverse ordering. This finding is also consistent with the results in \citet{ARRS21}.  But there are exceptions such as forecasting the Fed funds rate for $h=6$ and $h=12$. But the important point is not that, when using CS-VAR-SV models, one ordering is better than the other. The important points are that ordering matters (often in a statistically significant way) and that different variables prefer different orderings. That is, there is no one ordering that forecasts all of the variables best.  

\begin{table}[H]
	\centering	
	\caption{RMSFE and ALPL of four core macroeconomic time series.}
	\label{tab:forecasts}
\resizebox{\textwidth}{!}{\begin{tabular}{llcccccc}
\hline \hline
Variables             & Models      & \multicolumn{3}{c}{RMSFE}                                                             & \multicolumn{3}{c}{ALPL}                                                                                     \\ \cline{3-8} 
                      &             & $h=1$                            & $h=6$                   & $h=12$                   & $h=1$                             & $h=6$                              & $h=12$                              \\ \hline
Industrial production & CS-VAR-SV-1 & 0.007                            & \textbf{0.007}          & \textbf{0.007}           & 2.071                             & 2.159                              & 2.266                               \\
                      & CS-VAR-SV-2 & 0.007\ensuremath{^{**}}          & 0.007\ensuremath{^{**}} & 0.007\ensuremath{^{**}}  & 1.522\ensuremath{^{***}}          & 1.651\ensuremath{^{***}}           & 1.767\ensuremath{^{***}}            \\
                      & OI-VAR-SV-1 & \textbf{0.007}                   & 0.007                   & 0.007\ensuremath{^{***}} & \textbf{3.660}\ensuremath{^{***}} & 3.458\ensuremath{^{***}}           & \textbf{3.360}\ensuremath{^{***}}   \\
                      & OI-VAR-SV-2 & 0.007                            & 0.007                   & 0.007\ensuremath{^{***}} & 3.659\ensuremath{^{***}}          & \textbf{3.459}\ensuremath{^{***}}  & 3.358\ensuremath{^{***}}            \\ \cline{2-8} 
Unemployment rate     & CS-VAR-SV-1 & 0.159                            & 0.169                   & 0.173                    & $-$0.005                            & $-$0.689                             & $-$0.707                              \\
                      & CS-VAR-SV-2 & 0.158                            & 0.169                   & 0.173                    & $-$0.124                            & $-$0.761                             & $-$0.742                              \\
                      & OI-VAR-SV-1 & \textbf{0.158}                   & \textbf{0.167}          & \textbf{0.173}           & \textbf{0.463}\ensuremath{^{***}} & \textbf{0.276}                     & 0.152                               \\
                      & OI-VAR-SV-2 & 0.158                            & 0.167\ensuremath{^{*}}  & 0.173                    & 0.463\ensuremath{^{***}}          & 0.276                              & \textbf{0.156}                      \\ \cline{2-8} 
PCE inflation         & CS-VAR-SV-1 & 0.002                            & \textbf{0.002}          & \textbf{0.002}           & 2.210                             & 2.311                              & 2.425                               \\
                      & CS-VAR-SV-2 & 0.002\ensuremath{^{***}}         & 0.002                   & 0.002                    & 1.811\ensuremath{^{***}}          & 1.919\ensuremath{^{***}}           & 2.024\ensuremath{^{***}}            \\
                      & OI-VAR-SV-1 & \textbf{0.002}\ensuremath{^{**}} & 0.002                   & 0.002\ensuremath{^{*}}   & 4.821\ensuremath{^{***}}          & 4.448\ensuremath{^{***}}           & 4.249\ensuremath{^{***}}            \\
                      & OI-VAR-SV-2 & 0.002\ensuremath{^{**}}          & 0.002                   & 0.002                    & \textbf{4.823}\ensuremath{^{***}} & \textbf{4.453}\ensuremath{^{***}}  & \textbf{4.260}\ensuremath{^{***}}   \\ \cline{2-8} 
Federal funds rate        & CS-VAR-SV-1 & \textbf{0.492}                   & \textbf{1.566}          & \textbf{2.216}           & 0.222                             & $-$8.473                             & $-$16.704                             \\
                      & CS-VAR-SV-2 & 0.497                            & 1.597\ensuremath{^{*}}  & 2.264                    & $-$0.327\ensuremath{^{***}}         & \textbf{$-$6.708}\ensuremath{^{***}} & \textbf{$-$12.474}\ensuremath{^{***}} \\
                      & OI-VAR-SV-1 & 0.493                            & 1.576                   & 2.233                    & 0.294\ensuremath{^{***}}          & $-$6.956\ensuremath{^{*}}            & $-$13.466\ensuremath{^{*}}            \\
                      & OI-VAR-SV-2 & 0.494                            & 1.570                   & 2.236                    & \textbf{0.296}\ensuremath{^{***}} & $-$7.039                             & $-$13.662       
\\ \hline 
\end{tabular}}
{\raggedright 
\footnotesize Note: The bold figure indicates the best model in each case. \ensuremath{^{*}}, \ensuremath{^{**}} and \ensuremath{^{***}} denote, respectively, 0.10, 0.05 and 0.01 significance level for a two-sided \citet{DM95} test. The benchmark model is CS-VAR-SV-1.\par}
\end{table}

\section{Conclusions}

In this paper, we have demonstrated, both theoretically and empirically, the consequences of working with VAR-SVs which use Cholesky decompositions of the error covariance matrices such as that used in \citet{CS05} and are thus not order invariant. We have proposed a new specification which is order invariant which involves working with an unrestricted version of the impact matrix, $\bB_0$. Such a model would be unidentified in the homoscedastic VAR but we draw on \citet{BB20} to establish that the incorporation of SV identifies the model. We develop an MCMC algorithm which allows for Bayesian inference and prediction in our order-invariant model. In an empirical exercise involving $20$ macroeconomic variables we demonstrate the ability of our methods to produce accurate forecasts in a computationally efficient manner.

\newpage

\section*{Appendix A: Data Description}

This appendix provides details of the monthly dataset used in the forecasting exercise. The variables and their transformations are the same as in \citet{CCM19}.
 
\begin{table}[H]
	\centering
	\caption{Monthly dataset of 20 variables from FRED-MD.}
\begin{tabular}{lcc}
\hline \hline
Variable                                 & Mnemonic        & Transformation        \\ \hline
Real personal income                     & RPI             & $\Delta \text{log}$   \\
Real PCE                                 & DPCERA3M086SBEA & $\Delta \text{log}$   \\
Real manufacturing and trade sales       & CMRMTSPLx       & $\Delta \text{log}$   \\
Industrial production                    & INDPRO          & $\Delta \text{log}$   \\
Capacity utilization in manufacturing    & CUMFNS          & $\Delta$              \\
Civilian unemployment rate               & UNRATE          & $\Delta$              \\
Total nonfarm employment                 & PAYEMS          & $\Delta \text{log}$   \\
Hours worked: goods-producing            & CES0600000007   & no transformation     \\
Average hourly earnings: goods-producing & CES0600000008   & $\Delta \text{log}$   \\
PPI for finished goods                   & WPSFD49207      & $\Delta^2 \text{log}$ \\
PPI for commodities                      & PPICMM          & $\Delta^2 \text{log}$ \\
PCE price index                          & PCEPI           & $\Delta^2 \text{log}$ \\
Federal funds rate                       & FEDFUNDS        & no transformation     \\
Total housing starts                     & HOUST           & log                   \\
S\&P 500 price index                     & S\&P 500        & $\Delta \text{log}$   \\
U.S.-U.K. exchange rate                  & EXUSUKx         & $\Delta \text{log}$   \\
1 yr. Treasury-FEDFUNDS spread           & T1YFFM          & no transformation     \\
10 yr. Treasury-FEDFUNDS spread          & T10YFFM         & no transformation     \\
BAA-FEDFUNDS spread                      & BAAFFM          & no transformation     \\
ISM: new orders index                   & NAPMNOI        & no transformation \\ \hline \hline
\end{tabular}
\end{table}

\newpage 

\singlespace

\ifx\undefined\BySame
\newcommand{\BySame}{\leavevmode\rule[.5ex]{3em}{.5pt}\ }
\fi
\ifx\undefined\textsc
\newcommand{\textsc}[1]{{\sc #1}}
\newcommand{\emph}[1]{{\em #1\/}}
\let\tmpsmall\small
\renewcommand{\small}{\tmpsmall\sc}
\fi

\end{document}